\documentclass[twocolumn]{aastex631}
\usepackage{graphicx}

\newcommand\kms{\hbox{km$\,$s$^{-1}$}}
\newcommand\VLSR{\hbox{$V_{\rm LSR}$}}

\received{March 13, 2023}
\revised{March 31, 2023}
\accepted{March 31, 2023}
\submitjournal{ApJ}

\shorttitle{ALMA view of CO 0.02--0.02}
\shortauthors{Iwata et al.}

\graphicspath{{./}{figures/}}

\begin{document}

\title{ALMA View of the High-velocity-dispersion Compact Cloud CO 0.02--0.02 at the Galactic Center}

\correspondingauthor{Yuhei Iwata}
\email{yuhei.iwata@nao.ac.jp}

\author[0000-0002-9255-4742]{Yuhei Iwata}
\affiliation{Division of Science, National Astronomical Observatory of Japan, 2-21-1 Osawa, Mitaka, Tokyo 181-8588, Japan}
\affiliation{Center for Astronomy, Ibaraki University, 2-1-1 Bunkyo, Mito, Ibaraki 310-8512, Japan}
\affiliation{School of Fundamental Science and Technology, Graduate School of Science and
Technology, Keio University, 3-14-1 Hiyoshi, Kohoku-ku, Yokohama, Kanagawa 223-8522, Japan}

\author[0000-0002-5566-0634]{Tomoharu Oka}
\affiliation{School of Fundamental Science and Technology, Graduate School of Science and
Technology, Keio University, 3-14-1 Hiyoshi, Kohoku-ku, Yokohama, Kanagawa 223-8522, Japan}
\affiliation{Department of Physics, Faculty of Science and Technology, Keio University, 3-14-1
Hiyoshi, Kohoku-ku, Yokohama, Kanagawa 223-8522, Japan}

\author[0000-0001-8147-6817]{Shunya Takekawa}
\affiliation{Department of Applied Physics, Faculty of Engineering, Kanagawa University, 3-27-1 Rokkakubashi, Kanagawa-ku, Yokohama, Kanagawa 221-8686, Japan}

\author[0000-0002-1663-9103]{Shiho Tsujimoto}
\affiliation{Department of Physics, Faculty of Science and Technology, Keio University, 3-14-1
Hiyoshi, Kohoku-ku, Yokohama, Kanagawa 223-8522, Japan}

\author[0000-0003-2735-3239]{Rei Enokiya}
\affiliation{Department of Physics, Faculty of Science and Technology, Keio University, 3-14-1
Hiyoshi, Kohoku-ku, Yokohama, Kanagawa 223-8522, Japan}

\begin{abstract}
We report the results of observations toward the center of the molecular cloud CO 0.02--0.02 made using the Atacama Large Millimeter/Submillimeter Array. The successfully obtained $1\arcsec$ resolution images of CO {\it J}=3--2, H$^{13}$CN {\it J}=4--3, H$^{13}$CO$^{+}$ {\it J}=4--3, SiO {\it J}=8--7, CH$_3$OH {\it J}$_{\it K_a, K_c}$ = 7$_{1, 7}$--6$_{1, 6}$ A$^{+}$ lines, and 900 $\mu$m continuum show several new features, which have not been identified in previous observations. The dense gas probe (H$^{13}$CN, SiO, CH$_{3}$OH) images are dominated by a pair of northeast-southwest elongated filaments, which may be the main body of CO 0.02--0.02. Two striped patterns perpendicular to each other (F1 and F2) and a high-velocity feature (HV), which appear in different velocity ranges, were prominent in the CO image. An emission hole that may represent an expanding feature was found in the F1 velocity range. F2 appeared to align along the western edge of a $20\,\mbox{pc}\times 13\,\mbox{pc}$ ellipse (the Large Shell) identified in the single-dish CO map. The HV contains eight compact clumps at the positive high-velocity end of the CO emissions. Based on these results, we propose a formation scenario for CO 0.02--0.02; internal explosions of supernovae, external perturbations by the Large Shell, and gravitational acceleration by a less luminous star cluster have formed CO 0.02--0.02 in its current state.
\end{abstract}

\section{Introduction} 
\label{sec:intro}
The central few hundred pc region of our Galaxy contains a large amount of molecular gas and is referred to as the ``central molecular zone (CMZ)"~\citep{Morris96}. The molecular gas in the CMZ is higher in temperature and density than in the Galactic disk, showing highly complex distribution and kinematics. High-velocity-dispersion compact clouds are a peculiar population of molecular clouds found in the CMZ~\citep{Oka98,Oka22}. They are characterized by their compact appearance ($d\!<\! 5$ pc) and unusually large velocity widths ($\Delta V\! >\! 50$ \kms ). Although most of their driving sources are still unknown, several scenarios, such as supernova explosions from an embedded stellar cluster and gravitational kick by an inactive black hole, have been proposed as their origins so far~\citep[e.g.,][]{Oka01a,Oka16}. Because dense stellar clusters and black holes in central regions of galaxies are believed to be essential for the formation of supermassive black holes~\citep[e.g.,][]{Ebisuzaki01}, studies of these peculiar molecular clouds may contribute to the evolution of galactic nuclei.

\begin{figure*}[htbp]
\centering
\plotone{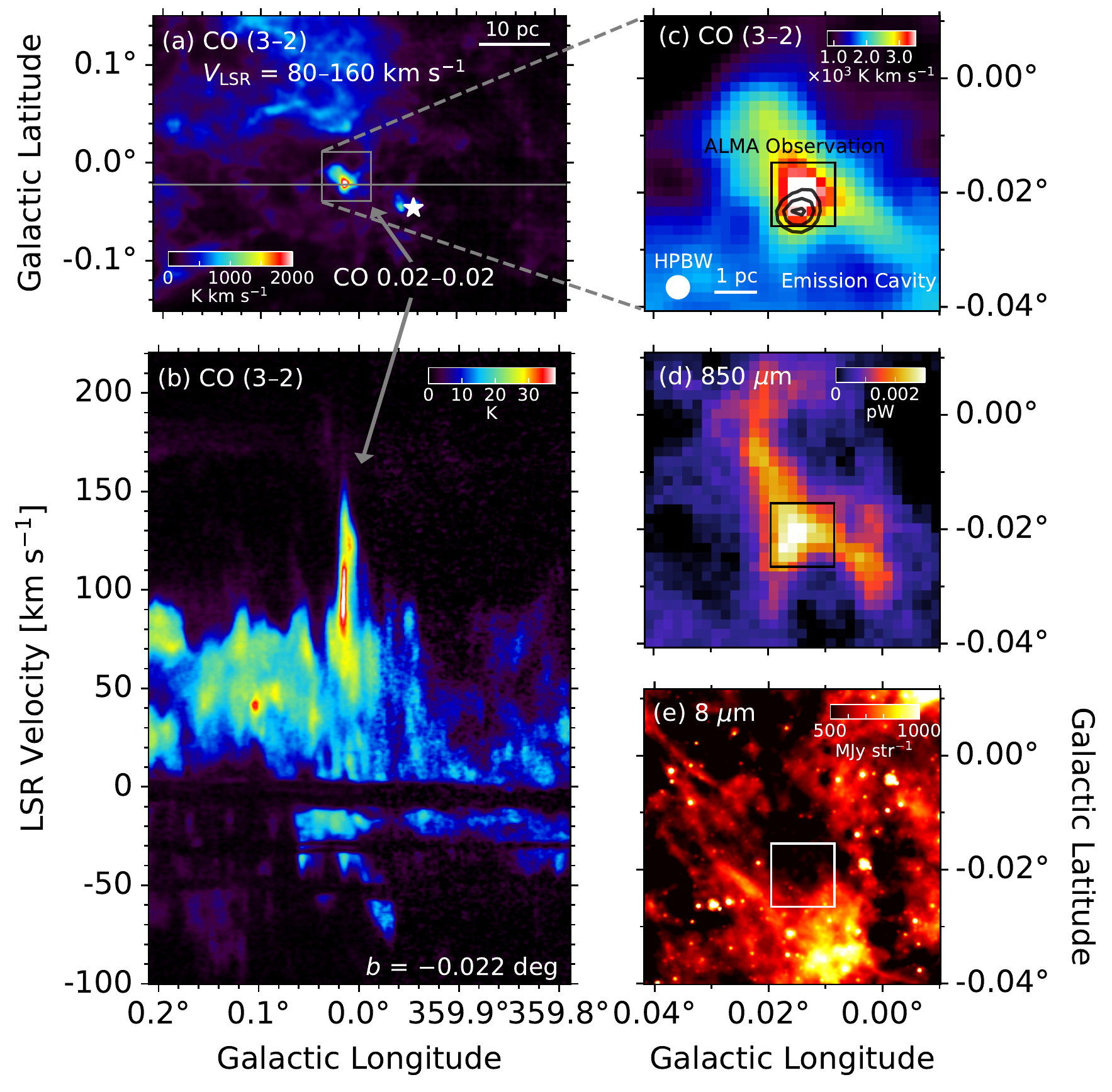}
\caption{(a) Integrated intensity map of CO {\it J}=3--2 line over $V_{\rm LSR}\! =\! 80$ to $160$ \kms\ obtained with the JCMT/HARP instrument~\citep{Parsons18,Eden20}. The white star denotes the location of Sgr A*. (b) Longitude-velocity map of CO {\it J}=3--2 line at $b\! =\! -0\fdg022$. (c) Zoomed map of the CO {\it J}=3--2 line over $V_{\rm LSR}$ = $0$ to $160$ \kms . The black contours show the distribution of the highest-velocity gas, integrated over $V_{\rm LSR}$ = $150$ to $160$ \kms. The black square indicates the mapping area of the ALMA Cycle 7 observations (same as the areas of the panels in Figure \ref{fig:integ}). (d) Continuum map at $850$ {\rm $\mu$m} obtained with the JCMT/SCUBA-2 instrument~\citep{Parsons18}. Contaminated CO {\it J}=3--2 emission is removed. (e) Spitzer/IRAC $8$ {\rm $\mu$m} map~\citep{Stolovy06}.
\label{fig:guide}}
\end{figure*}

CO 0.02--0.02 is one of the high-velocity-dispersion compact clouds located at $(l, b)\!\simeq\! (+0\fdg02, -0\fdg02)$, approximately 5\arcmin\ Galactic east (relative to the direction in the galactic coordinate reference frame) from Sgr A* (Figure \ref{fig:guide}a). This cloud stands out because of its extraordinary broad velocity width ($\Delta V\!\ge\! 100$ \kms , Figure \ref{fig:guide}b), compact size ($\simeq\!3\,\mbox{pc}\times 4\,\mbox{pc}$), and high CO {\it J}=3--2/{\it J}=1--0 intensity ratio \citep[$\simeq\! 1.3$;][]{Oka99}. The velocity at peak intensity $V_{\rm LSR}\!\simeq\! 100$ \kms\ was shifted by $\simeq\! 50$ \kms\ from the giant molecular cloud (GMC) M--0.02--0.07 (often referred to as ``the $+50$ \kms\ cloud''), which overlaps in the line-of-sight direction of CO 0.02--0.02. The total kinetic energy amounts to (3--8)$\times 10^{51}$ {\rm erg}~\citep{Oka99}, which corresponds to several tens of supernova explosions.

CO 0.02--0.02 is spatially elongated from northeast to southwest. An arc-shaped edge lies at the southwestern edge, surrounding the ``Emission Cavity''. Submillimeter and far-infrared surveys~\citep{Contreras13,Parsons18,Molinari16} have confirmed the dust-continuum counterpart of CO 0.02--0.02, which has a spatial structure similar to that of the molecular cloud (Figure \ref{fig:guide}d). No signature of massive star formation, such as \ion{H}{2} region or maser emission, was detected in CO 0.02--0.02. By inspecting mid-infrared images taken with the Spitzer Space Telescope~\citep{Stolovy06}, a group of point-like infrared sources was found within the Emission Cavity (Figure \ref{fig:guide}e). From these circumstances, CO 0.02--0.02 was suggested to be accelerated by supernova explosions, which formed the arc-shaped edge and Emission Cavity~\citep{Oka08}.

One of the remaining problems of the supernovae scenario is that the complicated kinematics of the cloud cannot be reproduced by coherent expanding motion. In particular, the positive highest-velocity gas of CO 0.02--0.02 at $V_{\rm LSR}\!\simeq\! 150$ \kms\ (contours in Figure \ref{fig:guide}c) shows inconsistencies with the supernovae-driven scenario. This component has a compact size of $\sim\! 14\arcsec$ and is located outside the Emission Cavity adjacent to the velocity-integrated emission peak. Since the edge of an expanding shell cannot accelerate the interstellar medium in the line-of-sight direction, this highest-velocity gas may not be the expanding origin. As the unresolved highest-velocity gas represents the compact and broad velocity dispersion peculiarity of CO 0.02--0.02, high-resolution observation is essential to understand its nature and origin.

To unveil the nature of CO 0.02--0.02, we conducted high-resolution and high-sensitivity observations using the Atacama Large Millimeter/submillimeter Array (ALMA). In this paper, we present the obtained high-angular-resolution images and reconsider the origin of the cloud based on the detailed kinematics of the highest-velocity gas. Throughout this paper, we assume that CO 0.02--0.02 is located at the distance to the Galactic center~\citep[$8.277$ {\rm kpc};][]{Gravity22}.

\section{Observations} \label{sec:obs}
Observations were performed as part of the ALMA Cycle 7 program (2019.1.01763.S; PI: Y. Iwata) using the $12$ {\rm m} array on 2019 December 5, $7$ {\rm m} array on 2019 December 10, 17, 20, and 2020 January 9, and the total power (TP) array on 2019 October 11, November 28, December 5, 7, 8, and 10. The observing area covers the central region of CO 0.02--0.02 and the highest-velocity gas (40\arcsec$\times$40\arcsec\ area, see the rectangle of Figure \ref{fig:guide}c), corresponding to the mosaic mapping of 27 pointings of the $12$ {\rm m} array and 10 pointings of the $7$ {\rm m} array. The $12$ {\rm m} array configuration was C43--1, with a baseline length of 15--161 {\rm m}. Four spectral windows (SPWs) in the two basebands were used to observe the molecular lines: CO {\it J}=3--2 (345.79 {\rm GHz}), H$^{13}$CN {\it J}=4--3 (345.34 {\rm GHz}), H$^{13}$CO$^{+}$ {\it J}=4--3 ($346.99$ {\rm GHz}), and SiO {\it J}=8--7 ($347.33$ {\rm GHz}). The two SPWs in the other two basebands with central frequencies of 333.0 and $334.9$ {\rm GHz} were used to observe the CH$_3$OH {\it J}$_{\it K_a, K_c}$ = 7$_{1, 7}$--6$_{1,6}$ A$^{+}$ line ($335.58$ {\rm GHz}) and continuum emissions. All six SPWs were observed simultaneously. The bandwidths were $937.5$ {\rm MHz} for the first four SPWs and $1875.0$ {\rm MHz} for the latter two SPWs. The frequency resolution of all SPWs was $976.562$ {\rm kHz}. Quasars J1924--2914 and J1517--2422 were observed as flux and bandpass calibrators. The phase calibrators used were J1744--3116 and J1717--3342. Although the status of the initial quality assurance (QA0) of the $12$ {\rm m} array data was ``semipass'', we decided to use the data because there seemed to be no severe problems. The data of the other two arrays were QA2-passed. 

We calibrated and reduced the data using the Common Astronomy Software Applications (CASA) 5.7 \citep{McMullin07,CASA22}. Calibrated data were generated using the calibration script provided by the East Asian ALMA Regional Center. Continuum emission was subtracted from the spectral data using the task ``uvcontsub''. The interferometric images were created using the task ``tclean'' with Briggs weighting and a robust parameter of 0.5, while the TP images were created using ``sdimaging''. Each interferometric image and TP image were combined with the task ``feather''. The spatial and velocity grid widths of the resultant images were 0.1\arcsec\ and $2$ \kms, respectively. The synthesized beam size of the combined spectral cubes was $1\farcs2 \times 0\farcs9$ with a position angle of $-22\arcdeg$. The standard deviation of the cubes was $10$ {\rm mJy beam$^{-1}$}.

\begin{figure*}[hbtp]
\centering
\plotone{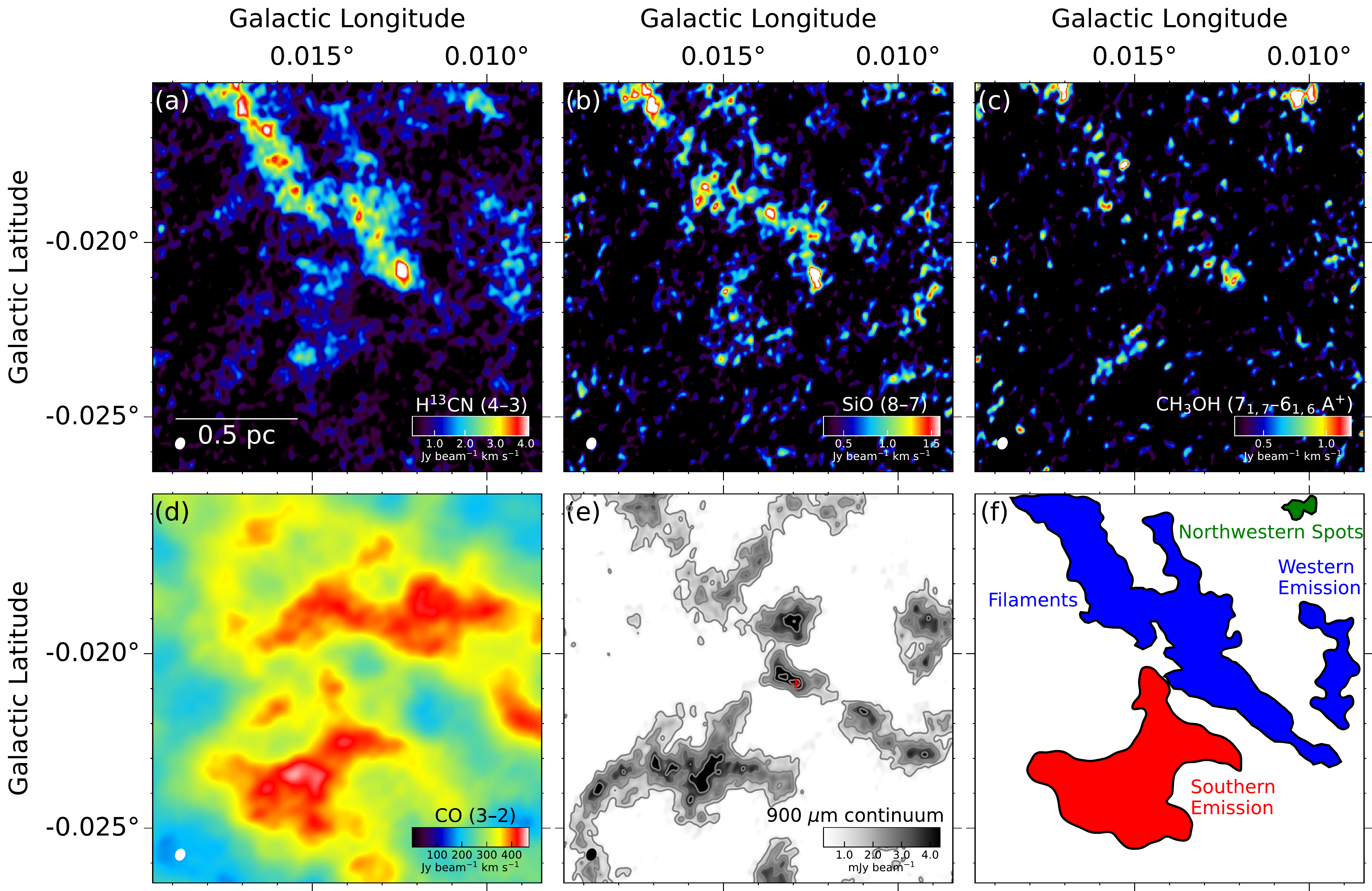}
\caption{(a--d) Integrated intensity maps of the H$^{13}$CN, SiO, CH$_3$OH, and CO lines over $V_{\rm LSR}$ = $-50$ to $+250$ \kms. The velocity range is set to avoid contamination of emissions from other spectral lines. The synthesized beam is shown as a white ellipse at the bottom left of each panel. The maps are produced by combining the interferometric (12 m and 7 m) images with the TP images. (e) The interferometric image of $900$ {\rm $\mu$m} continuum emission. The contour levels are 1, 2, 3, 4, and $5$ {\rm mJy beam$^{-1}$}. The standard deviation is $1.6$ {\rm mJy beam$^{-1}$}.  The red cross indicates the position of the brightest continuum core. (f) The schematic view of the characteristic features. Colors represent spectral lines in which each feature is prominent; blue: H$^{13}$CN, green: CH$_3$OH, and red: CO. The Filaments are more elongated toward the southwest than in the integration map, taking into account the velocity channel map in Figure \ref{fig:h13cn}. 
\label{fig:integ}}
\end{figure*}

\section{Results}
\subsection{Spatial Distributions}
Figure \ref{fig:integ} shows the velocity-integrated maps of the four observed molecular line emissions (H$^{13}$CN, SiO, CH$_3$OH, and CO), the interferometric image of $900$ {\rm $\mu$m} continuum emission, and schematic view of the characteristic features in the integrated maps. Note that the H$^{13}$CO$^{+}$ emission was not detected in the integrated map. The spatial distributions of the H$^{13}$CN, SiO, and CH$_3$OH line emissions were roughly similar. We clearly see two filaments of intense H$^{13}$CN emissions which range northeast to southwest.  We hereafter refer to these as the ``Filaments''. The faint SiO and CH$_3$OH emissions trace the Filaments, except for the CH$_3$OH emission spots near the northwest corner. These northwestern CH$_3$OH spots were also visible in the H$^{13}$CN map. We also observed faint H$^{13}$CN emission features in the south and west of the Filaments. Significantly faint SiO and CH$_3$OH emissions appear to be associated with the southern and western H$^{13}$CN features.

The spatial distribution of the CO line emission (Figure \ref{fig:integ}d) differs completely from that of the other observed molecular lines. The wide spatial extent of the CO emissions may be due to the lower critical density for excitation ($n_{\rm cr}\!\simeq\! 10^{4.5}$ cm$^{-3}$) and ubiquity of the CO molecule. We carefully checked whether the CO image was recovered by comparing the smoothed ALMA image with the JCMT image. The Filaments, as well as the western H$^{13}$CN feature and the northwestern CH$_3$OH spots, seem to be buried in the spatially extended CO emission. The southern H$^{13}$CN feature corresponds to the center of a more spatially extended feature of CO emissions, where the CO integrated intensity is the most intense.

The $900$ {\rm $\mu$m} continuum image (Figure \ref{fig:integ}e) shows the characteristics that are a mixture of those of H$^{13}$CN and CO. Continuum emission features trace the Filaments, further elongating them toward southwest. The western H$^{13}$CN emission region is also visible in the continuum. In addition, an intense, spatially extended continuum emission feature was observed in the southern part. The Filaments and the southern emission feature trace the large-scale molecular gas distribution observed with single-dish telescopes (see Figure \ref{fig:guide}c).

\begin{figure*}[htbp]
\centering
\plotone{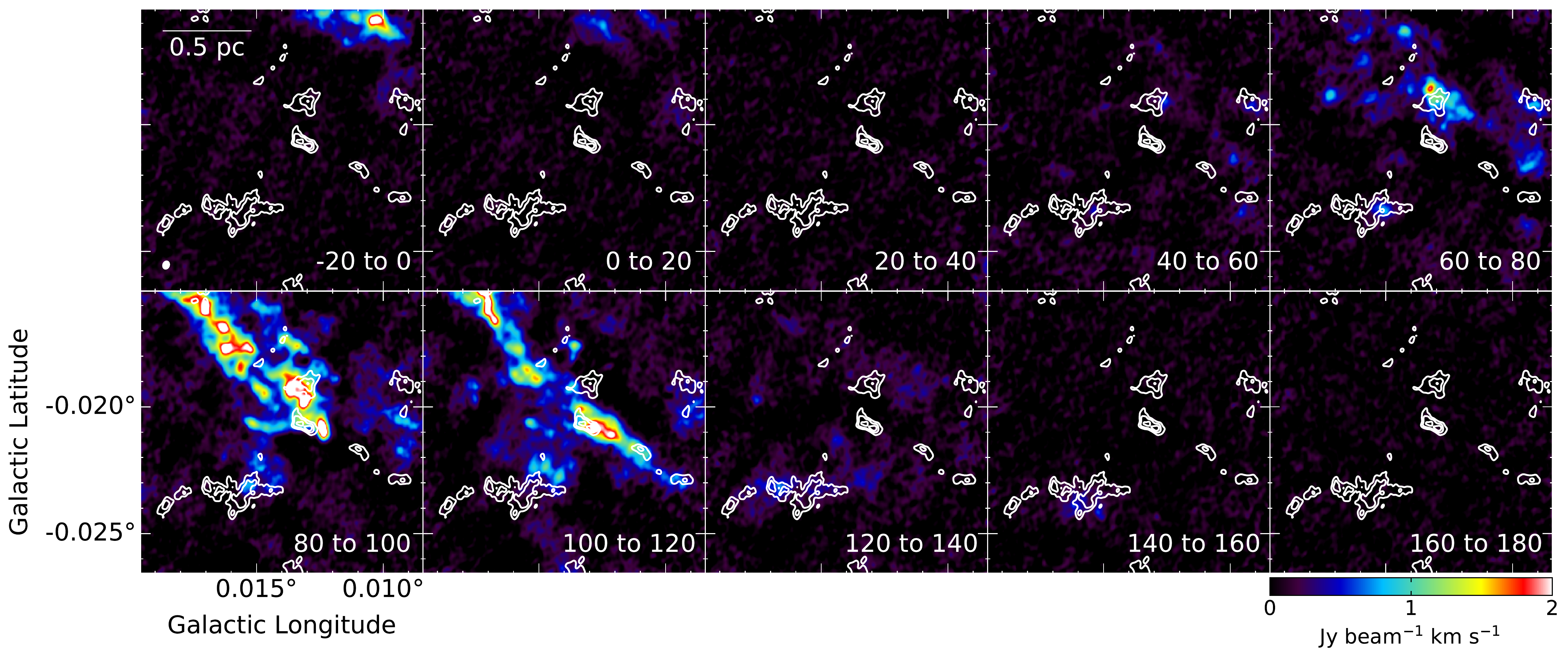}
\caption{Velocity channel maps of H$^{13}$CN {\it J}=4--3 line emission. The integrated velocity range is indicated at the bottom right of each panel. The synthesized beam is shown as a white ellipse in the top left panel. The white contours show the $900$ {\rm $\mu$m} continuum emission (Figure \ref{fig:integ}e). The contour levels are 3, 4, and $5$ {\rm mJy beam$^{-1}$}.
\label{fig:h13cn}}
\end{figure*}

\subsection{Velocity Structures}
\label{subsec:vel_st}
Figure \ref{fig:h13cn} shows velocity channel maps of the H$^{13}$CN line. The Filaments appeared in the velocity range between $\VLSR\! =\! 60$ and $120$ \kms , with a velocity gradient from northeast to southwest. The southern feature ranges from $\VLSR\! =\! 60$ to $160$ \kms , breaking up into several clumps of H$^{13}$CN emissions. The western feature appears at $\VLSR\! =\! 60$ to $120$ \kms\ and breaks up into a few clumps. The northwestern CH$_3$OH spots appeared at $\VLSR\! =\! -20$ to $0$ \kms , which suggests a possible irrelevance to CO 0.02--0.02.    

\begin{figure*}[htbp]
\centering
\plotone{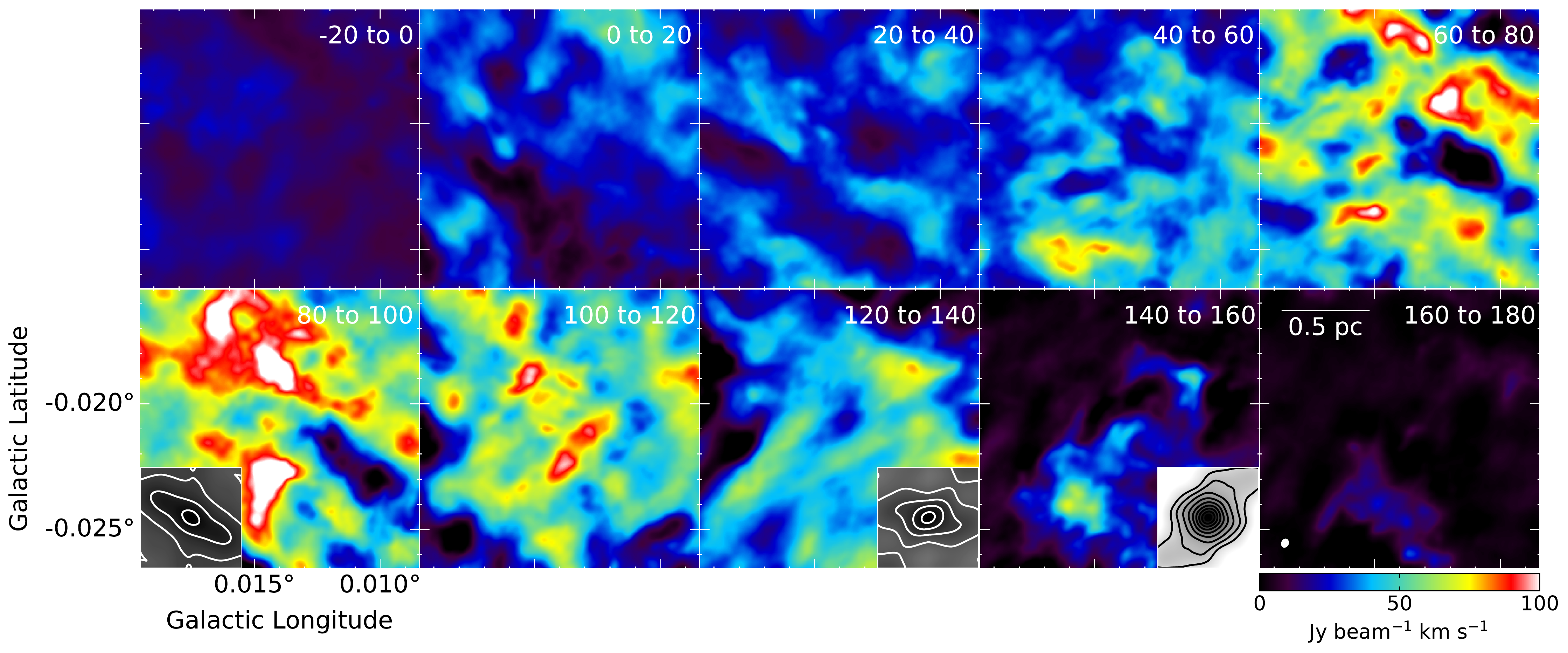}
\caption{Velocity channel maps of CO {\it J}=3--2 line emission. The velocity ranges are the same as Figure \ref{fig:h13cn}. The synthesized beam is shown as a white ellipse in the lower right panel. The autocorrelation maps, calculated by handling the boundaries by filling zero and normalizing each pixel value, are shown in the three panels. The highest contour level is 0.98 and the intervals are 0.05. The plotting region of each autocorrelation map corresponds to the spatial scale of 65\% region of the ALMA map. The three autocorrelation maps represent the characteristics of F1, F2, and HV. \label{fig:co_channel}}
\end{figure*}

\begin{figure}[htbp]
\centering
\plotone{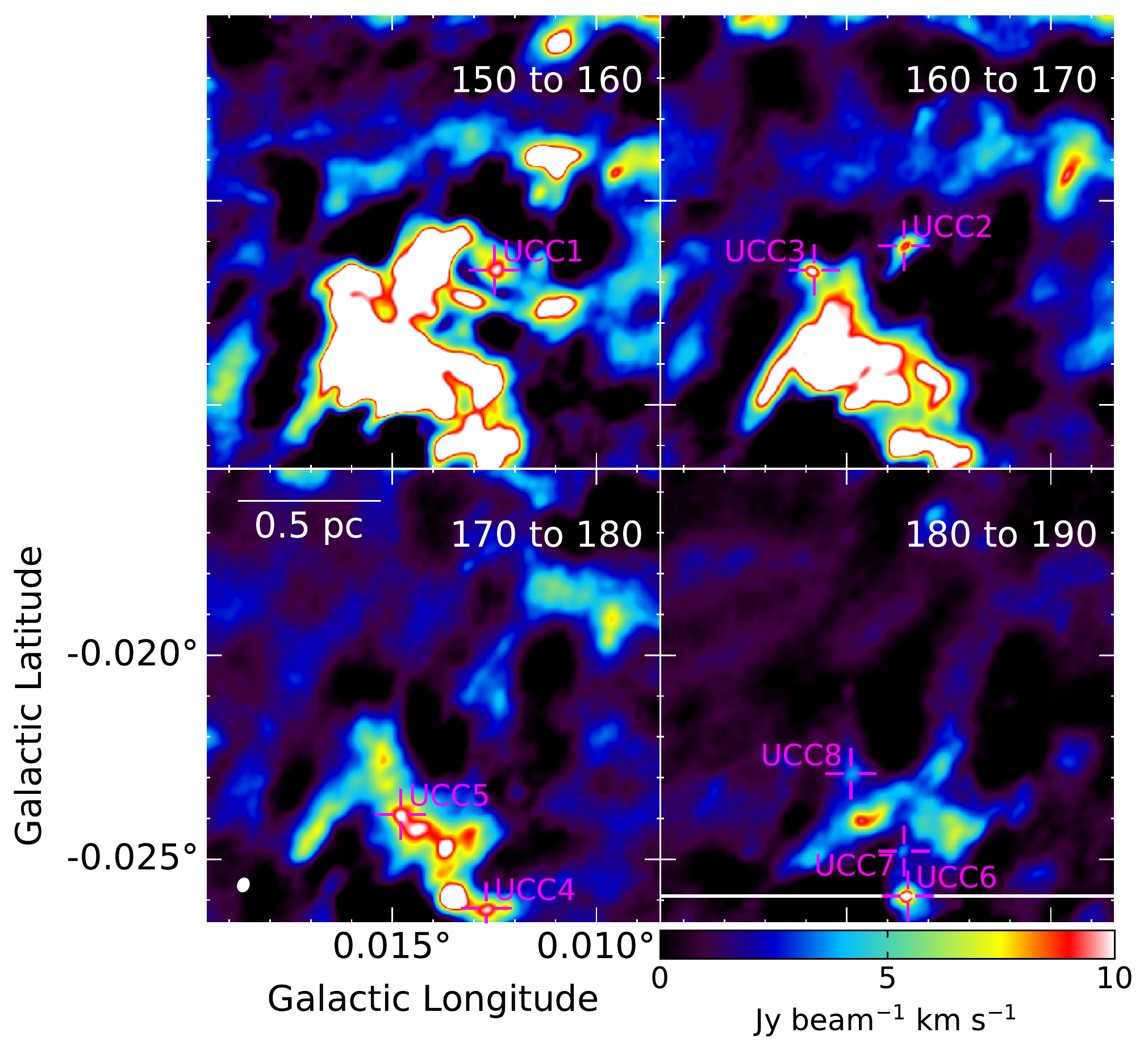}
\caption{Velocity channel maps of the CO {\it J}=3--2 line from $V_{\rm LSR} = 150$ to $190$ \kms , representing the HV. The integrated velocity range is indicated at the top right of each panel. The synthesized beam is shown as a white filled ellipse in the lower left panel. Magenta crosses indicate the positions of the ultra-compact clumps. The white horizontal line in the lower right panel corresponds to the slice of Figure \ref{fig:p-v}.
\label{fig:chv_channel}}
\end{figure}

\begin{figure}[ht!]
\centering
\plotone{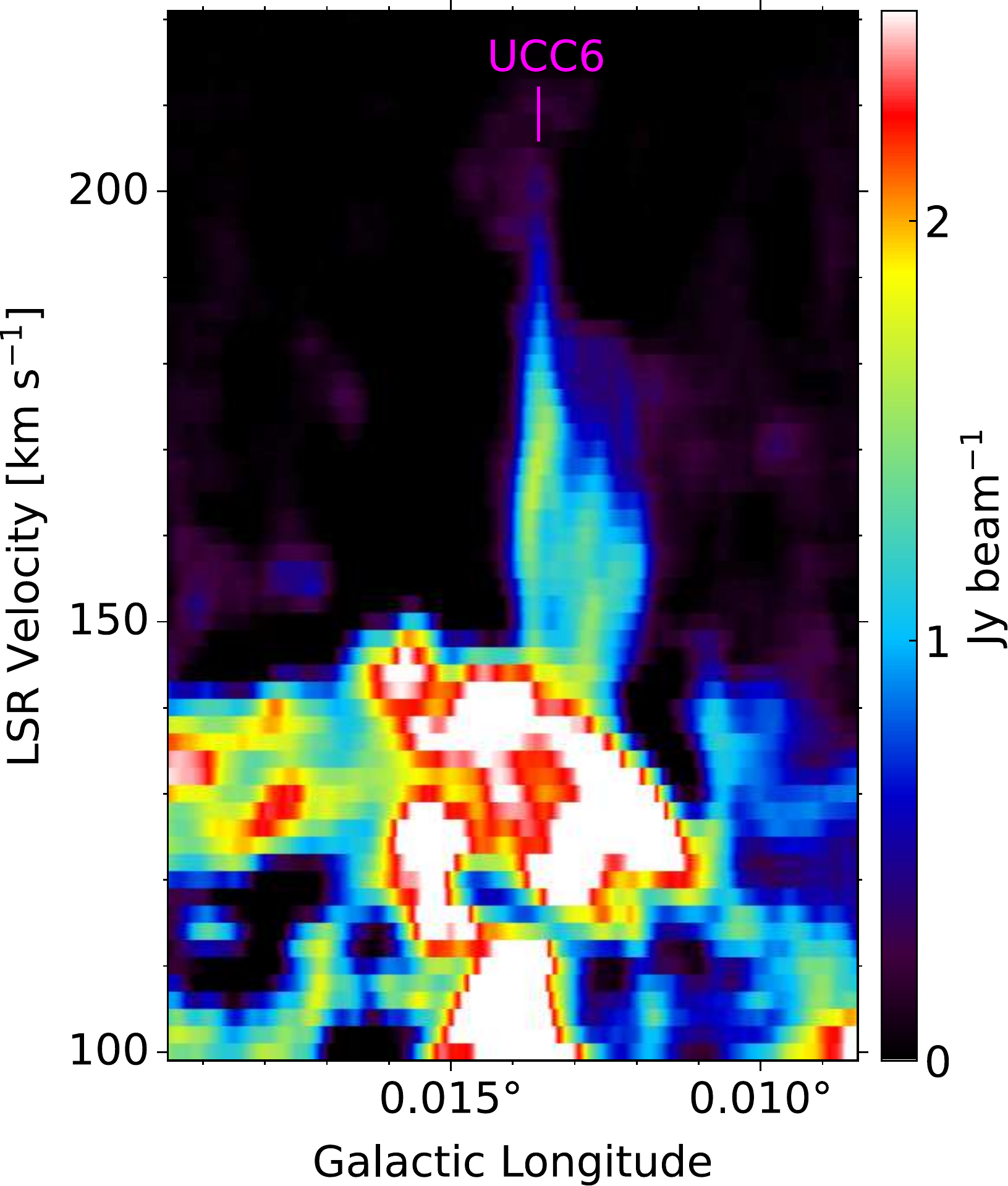}
\caption{Longitude-velocity map of the CO {\it J}=3--2 line at $b = -0\fdg 026$ (the white horizontal line in Figure \ref{fig:co_channel}).
\label{fig:p-v}}
\end{figure}

Figure \ref{fig:co_channel} shows the velocity channel maps of the CO emissions. At the slight west of the center in the $\VLSR\!=\!60\mbox{--}80$ and $80\mbox{--}100$ \kms\ channel maps, there is a hole of emission. Its direction is similar to that of the Filaments. We hereafter refer to the hole as ``Slit". The Slit also can be seen in the CO integrated intensity map of Figure \ref{fig:integ}d.

The wide spatial extent of CO emissions hinders our understanding of its spatial velocity structure, especially at velocities lower than $\VLSR\! =\! 140$ \kms . Upon investigating the CO velocity channel maps minutely, we noticed two distinctive filamentary features (F1 and F2) and a high-velocity component (HV). These features appearing at different line-of-sight velocity ranges are explained in detail below.

The distribution of CO emissions at velocities between $\VLSR\! =\! 0$ and $100$ \kms\ shows a rough striped pattern in the northeast-southwest direction. To characterize such a pattern, we calculated autocorrelations of the CO channel maps and show characteristic three maps in the subpanels inserted in Figure \ref{fig:co_channel}. The correlated region corresponds to the spatial scale of the clouds, and elongation reflects the typical cloud shape and direction. The northeast-southwest direction pattern is manifested in the autocorrelation maps of $\VLSR\! =\! 80$ to $100$ \kms . Hereafter, we refer to this northeast-southwest striped emission feature as ``F1".  The aforementioned two Filaments align along F1. At velocities between $\VLSR\! =\! 100$ and $140$ \kms , another rough striped pattern of the southeast-northwest direction appears. This is hereafter referred to as ``F2".  The mesh-patterned CO emission features at the $\VLSR\! =\! 100\mbox{--}120$ velocity channel is a mixture of F1 and F2. In contrast to F1, we could not detect any H$^{13}$CN, SiO, CH$_3$OH, or H$^{13}$CO$^{+}$ line emissions from F2. The absence of these high-density probe emissions indicates that less dense gas dominates F2.

Another distinct feature appeared at velocities higher than $\VLSR\! =\! 140$ \kms , as a $\simeq\! 7\arcsec$ diameter CO emission, at the most intense part of the southern continuum emission feature. Hereafter, we referred to this high-velocity feature as ``HV". The HV corresponds to the positive highest-velocity gas identified in the JCMT CO {\it J}=3--2 map (the contours in Figure \ref{fig:guide}c) as well as to the positive high-velocity end of the southern emission feature in the velocity-integrated maps (see Figure \ref{fig:integ}f). F2 and HV seem to be connected in the position-velocity structure because the shape of the HV shrinks at the higher velocity range from F2. The autocorrelation map reflects the spatial compactness of the HV (Figure \ref{fig:co_channel}).  

Figure \ref{fig:chv_channel} shows CO velocity channel maps between $\VLSR\!=150$ \kms\ and $190$ \kms\ with a 10 \kms\ interval. We found eight ultra-compact clumps (UCCs), which appear as isolated compact shapes at their positive high-velocity ends, and summarized them in Table \ref{table:ucc}. These UCCs have angular sizes that are approximately like those of the synthesized beam, indicating they these may be unresolved. Such ultra-compact clumps were only detected in the angular area of the HV envelope. UCC6 is visible in the CO {\it J}=3--2 position-velocity map (Figure \ref{fig:p-v}) up to $V_{\rm LSR}\!\simeq\! 200$ \kms. This is the ``real" highest-velocity gas in CO 0.02--0.02.

\begin{deluxetable}{cccc}
\tablecaption{Ultra-compact clumps in CO 0.02--0.02\label{table:ucc}}
\tablehead{
\colhead{ID} & \colhead{$l$} & \colhead{$b$} & \colhead{$V_{\rm LSR}$\tablenotemark{a}}\\
\colhead{} & \colhead{(deg)} & \colhead{(deg)} & \colhead{(\kms )}
}
\startdata
UCC1 & $0.0125$ & $-0.0217$ & 150 to 160 \\
UCC2 & $0.0136$ & $-0.0211$ & 160 to 170 \\
UCC3 & $0.0158$ & $-0.0217$ & 160 to 170 \\
UCC4 & $0.0127$ & $-0.0262$ & 170 to 180 \\
UCC5 & $0.0148$ & $-0.0239$ & 170 to 180 \\
UCC6 & $0.0135$ & $-0.0259$ & 180 to 190 \\
UCC7 & $0.0136$ & $-0.0248$ & 180 to 190 \\
UCC8 & $0.0149$ & $-0.0229$ & 180 to 190 \\
\enddata
\tablenotetext{a}{The LSR velocity is taken from Figure \ref{fig:chv_channel}.}
\end{deluxetable}

\begin{figure}[ht!]
\plotone{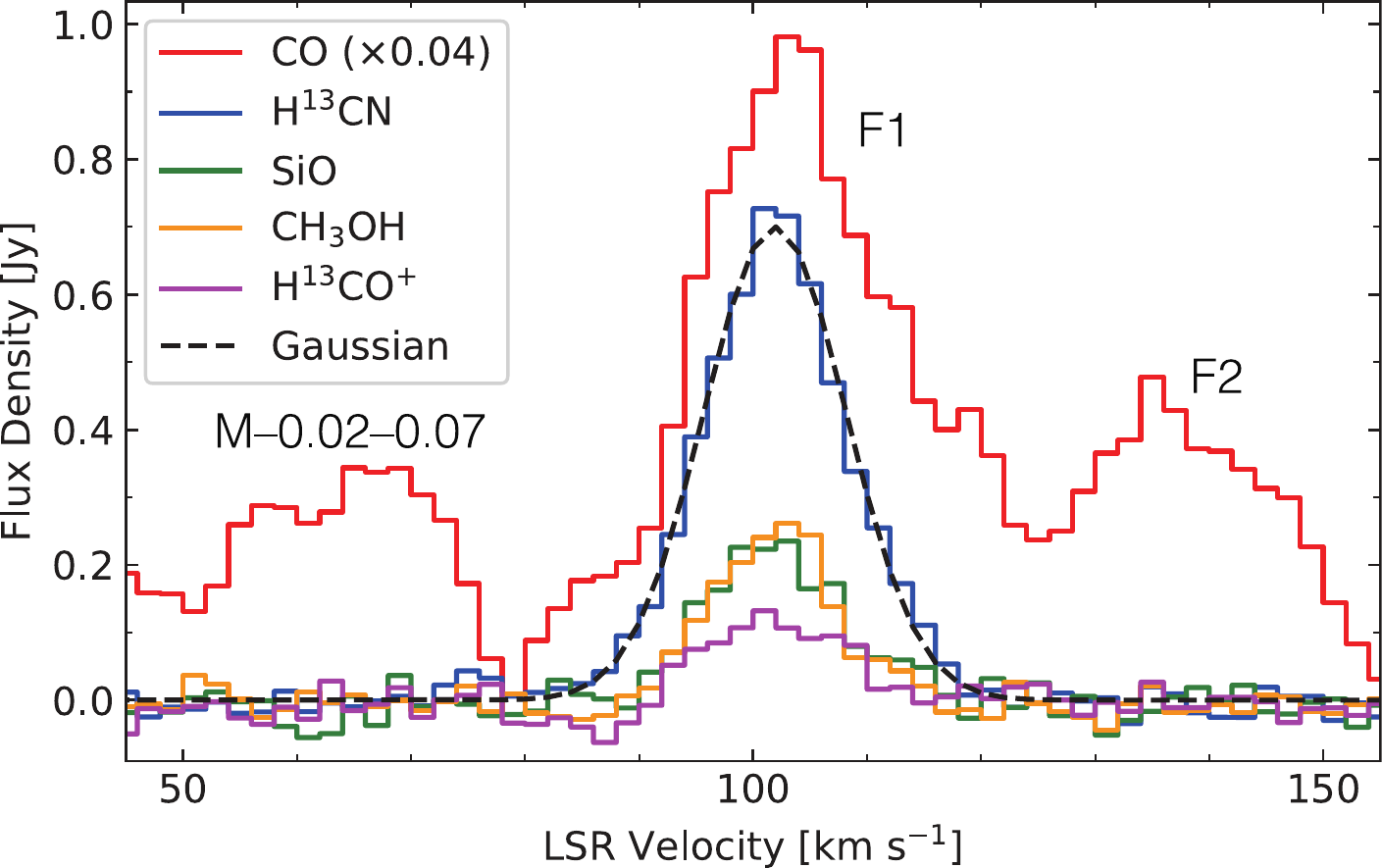}
\caption{Spectra of the CO, H$^{13}$CN, SiO, CH$_3$OH, H$^{13}$CO$^{+}$ lines at ($l, b$) = ($0\fdg 0129, -0\fdg 0209$). The CO spectrum is multiplied by a factor of 0.04 for the sake of clarity. A Gaussian curve fitted to the H$^{13}$CN spectrum is shown as a black dashed line.
\label{fig:spectrum}}
\end{figure}

\subsection{Brightest Continuum Core}
We observe several small $900$ {\rm $\mu$m} continuum cores along the Filaments and the southern emission feature (Figure \ref{fig:integ}e). The brightest continuum emission originates from $(l, b)\! =\! (0\fdg 0129, -0\fdg 0209)$, where a $\simeq\! 2\arcsec$-sized core is located (Figure \ref{fig:integ}e). Figure \ref{fig:spectrum} shows the molecular line spectra extracted from this brightest continuum core (BCC) region. All observed molecular lines were detected, including H$^{13}$CO$^{+}$ with a 4.8$\sigma$ level at the peak. All spectra, except for the CO line, had similar shapes. The CO spectrum consisted of three components. Two of these are attributed to the F1 and F2 features, and the lower velocity component at approximately $\VLSR\! =\! 50$ \kms\ arises from M--0.02--0.07.

By fitting a single Gaussian function to the H$^{13}$CN spectrum, we obtained the velocity dispersion of $\sigma_{V}\! =\! 6.3$ \kms\ with a center velocity of $101.9$ \kms . The H$^{13}$CN/H$^{13}$CO$^{+}$ ratio at the H$^{13}$CN peak velocity was 5.5, which is a normal value in the CMZ \citep[2.6--6.6;][]{Jones12}. The size parameter ($S\!\equiv\! D\mathrm{tan}\sqrt{\sigma_{l}\sigma_{b}}$) of the BCC was estimated to be $0.05$ pc using H$^{13}$CN data. At $S\! =\! 0.05$  pc, the velocity dispersion of $\sigma_{V}\! =\! 6.3$ \kms\ is approximately one order of magnitude higher than that estimated from the extrapolated $S$-$\sigma_{V}$ relation of the CMZ clouds \citep{Oka01b,Shetty12}. This behavior follows that of the high velocity-dispersion compact clouds \citep{Tokuyama17,Oka22}.

\begin{figure*}[htbp]
\centering
\plotone{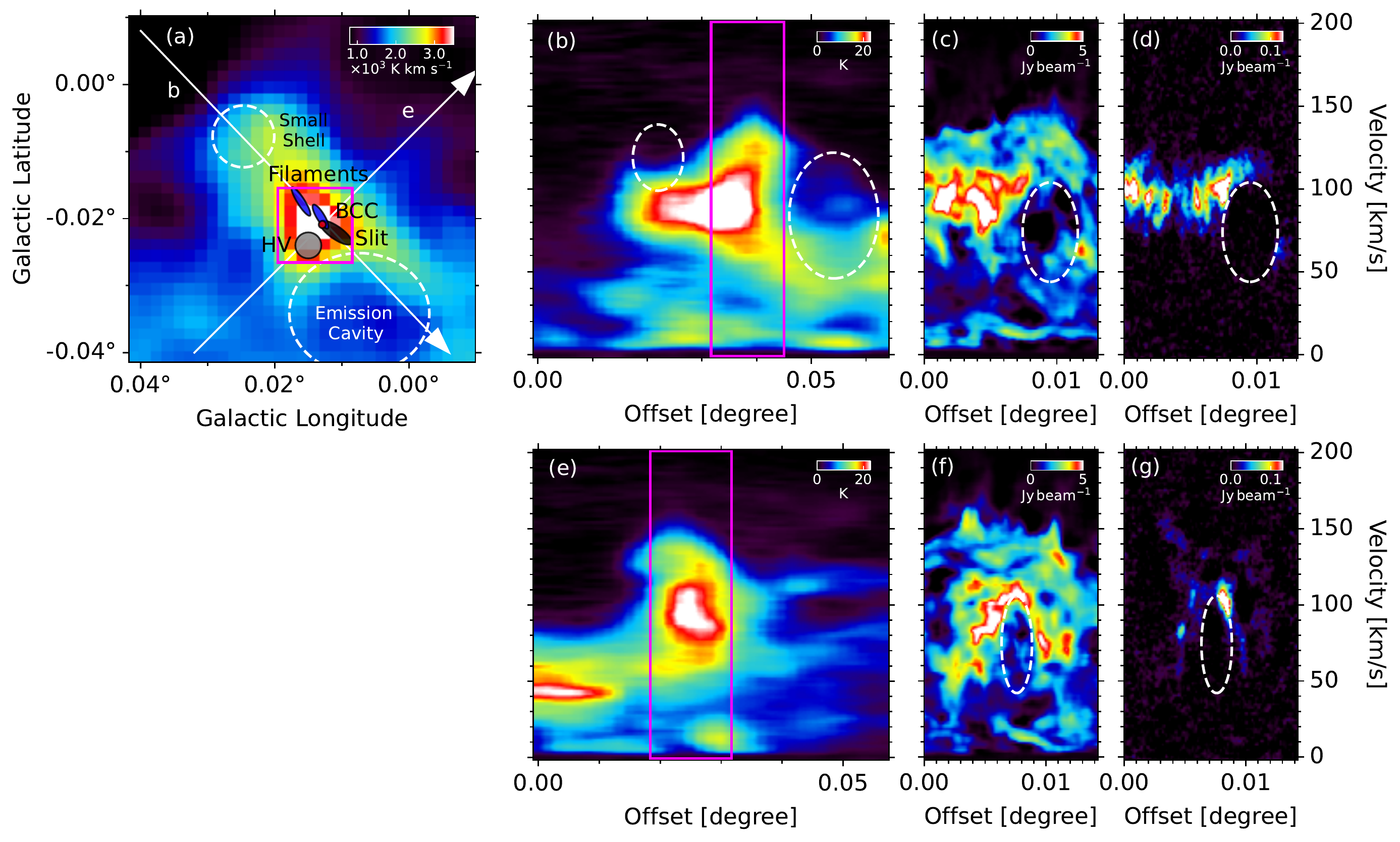}
\caption{(a) Schematic view of the identified features in CO 0.02--0.02 superposed on CO {\it J}=3--2 integrated intensity map obtained with the JCMT. Two white arrows ``b" and ``e" denote the positional cuts where the position-velocity diagrams are produced. The white dashed ellipses show the Emission Cavity and the Small Shell indicated in \citet{Oka08}. (b) Position-velocity diagram along the arrow ``b" from the JCMT CO {\it J}=3--2 data. The data were averaged over a $6\arcsec$ width vertical to the positional cut. The white dashed ellipses correspond to the position-velocity structures of the Emission Cavity and the Small Shell. (c)--(d) Position velocity diagrams from the ALMA CO {\it J}=3--2 and H$^{13}$CN {\it J}=4--3 lines at the magenta rectangle in the panel (b). The data were averaged over a $1\farcs 5$ width vertical to the positional cut. (e) Position-velocity diagram along the arrow ``e" from the JCMT CO {\it J}=3--2 line. The data were averaged over a $6\arcsec$ width vertical to the positional cut. (f)--(g) Position velocity diagrams from the ALMA CO {\it J}=3--2 and H$^{13}$CN {\it J}=4--3 data at the magenta rectangle in the panel (e). The data were averaged over a $1\farcs 5$ width vertical to the positional cut. The white dashed ellipses in (c)--(d) and (f)--(g) indicate the position-velocity structures of the Slit.
\label{fig:endogen}}
\end{figure*}

\section{Discussion}
ALMA observations of CO 0.02--0.02 have revealed its spatial fine structures (Slit, BCC, F1, F2, HV, and UCCs) that were previously unrecognized. The Filaments, which are a part of F1, seem to be the spine of CO 0.02--0.02. This suggests that the main body of CO 0.02--0.02 is centered at $\VLSR\!\simeq\! 100$ \kms . F2, having a different alignment direction from F1, suggests an interaction with another component with $\VLSR\!\simeq\! 120$ \kms . However, the localized features of Slit, HV, and UCCs may indicate acceleration through the internal injection of kinetic energy. Here, we discuss the origin of CO 0.02--0.02 by synthesizing the results of observations made so far.

\subsection{Endogenesis}
The originally proposed formation scenario of CO 0.02--0.02 is based on the acceleration by a series of supernova explosions~\citep{Oka99, Oka08}.  This idea came from the association of the ``Small Shell'' and ``Emission Cavity'' with a group of point-like infrared sources (stellar cluster candidate). This scenario was also successful in energetics, requiring a young ($\mbox{age}\! =\! 10\mbox{--}30$ Myr) and very massive ($M\!\sim\! 10^{6}$ $M_{\odot}$) star cluster in the immediate vicinity of CO 0.02--0.02~\citep{Oka08}.   

Figure \ref{fig:endogen}a summarizes the characteristic features in CO 0.02--0.02 identified in the single-dish data~\citep{Oka99, Oka08} and the ALMA data. The Slit, BCC, and HV were spatially adjacent to the Emission Cavity, which included the stellar cluster candidate. The CO {\it J}=3--2 position velocity maps along and perpendicular to the Slit (Figure \ref{fig:endogen}c) show that the Slit closed at $\VLSR\!\simeq\! 110$ \kms , tracing a rough half-ellipsoid in the position-velocity space. This behavior suggests the expansion of the kinematics of the Slit. The southwestern tip of the Filaments, where the BCC was located, also traced this ellipsoid (Figure \ref{fig:endogen}f). The broad velocity width of the BCC may be partially owing to its interaction with the Slit.  

The central velocity of the Slit ellipsoid $\VLSR\!\simeq\! 75$ \kms , was almost equal that of the Emission Cavity. The proximity in position and velocity suggests that their origins are closely related. The expansion time of the Slit is estimated to be $\sim\! 10^{4}$ yr, a factor of three shorter than that of the Emission Cavity. This discrepancy suggests that a series of explosions, which are responsible for the formation of the Emission Cavity, still continues in CO 0.02--0.02. The discovery of the Slit may reinforce the endogenous acceleration of CO 0.02--0.02, but it cannot account for the higher-velocity features, that are F2 and HV.   

\begin{figure}[ht!]
\plotone{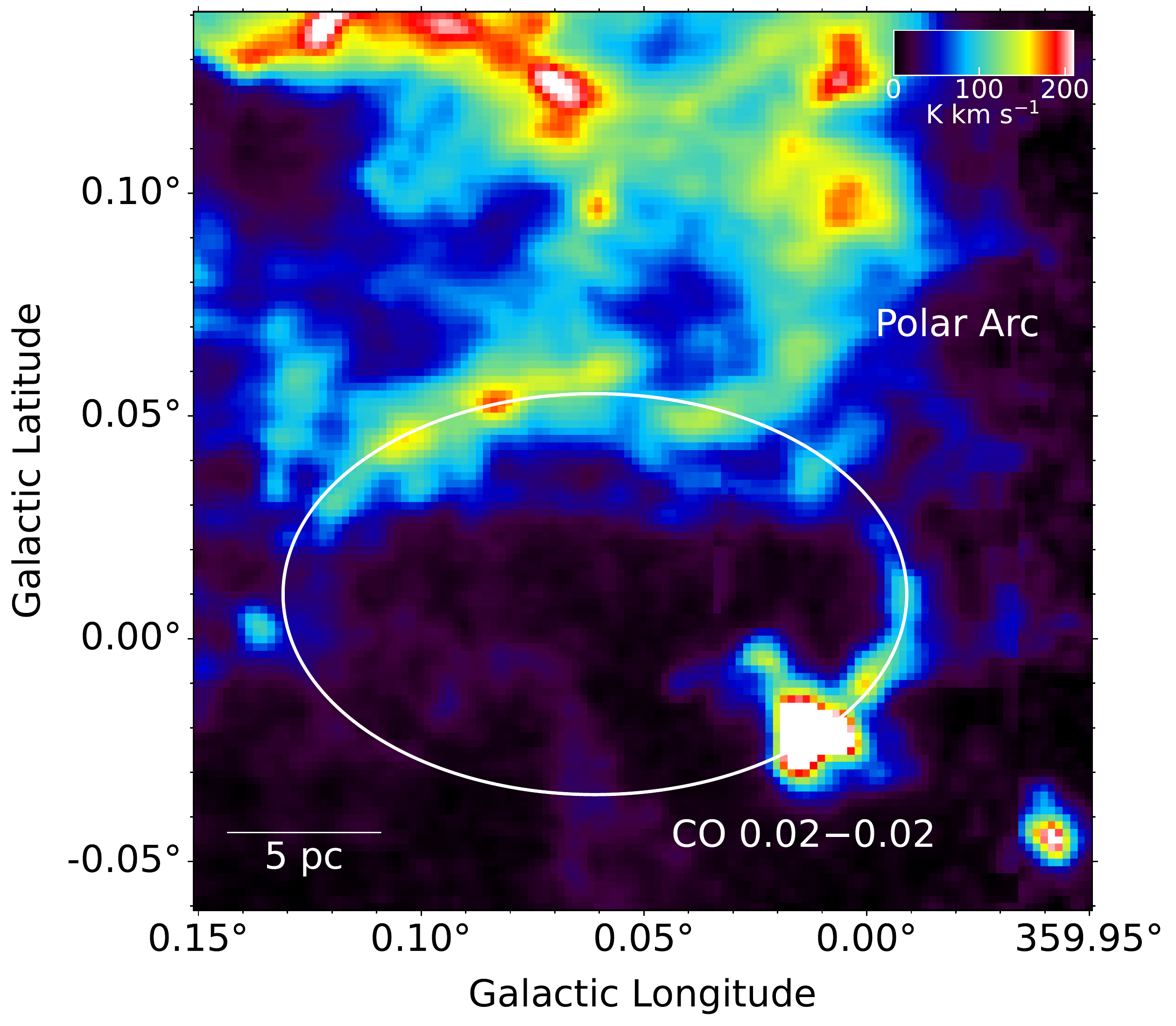}
\caption{Integrated intensity map of the CO {\it J}=3--2 line obtained with JCMT over $V_{\rm LSR}$ = $110$ to $130$ \kms. The white ellipse represents the Large Shell.
\label{fig:co32_shell}}
\end{figure}

\subsection{Exogenesis}
The spatial structure of CO 0.02--0.02 changes at $\VLSR\!\simeq\! 120$ \kms\ from F1 to F2. Because the HV seems to arise from F2, their origins are likely to be closely related. Taking another look at the full view of CO 0.02--0.02 (Figure \ref{fig:guide}c), we noticed a northwestern protrusion extending from its intensity peak, along the direction of F2. This protrusion appears only in the CO maps \citep{Oka99} at $\VLSR\!\simeq\! 120$ \kms (Figure \ref{fig:endogen}e), having similar chemical composition of F2 (Figure \ref{fig:spectrum}). These similarities may suggest a physical association of protrusion with F2.

In the large-scale view (Figure \ref{fig:guide}a), the protrusion curves toward the north appear to connect to the southern end of the Polar Arc \citep{Bally88}. Figure \ref{fig:co32_shell} shows that the northwestern protrusion is part of a $20\,\mbox{pc}\times\! 13\,\mbox{pc}$-sized ellipse at the bottom of the Polar Arc. The ellipse appearing at $\VLSR\!=\! 100\mbox{--}140$ \kms\ could be a manifestation of a large-scale expanding shell (hereafter referred to as ``Large Shell''). The lack of an emission depression at $\VLSR\!\leq\! 80$ \kms\ suggests that the expanding velocity and expansion time of the Large Shell are likely to be $\simeq\! 40$ \kms\ and $\simeq\! 10^{5.5}$ yr, respectively.

The presence of a Large Shell may evoke an exogenous acceleration scenario for CO 0.02--0.02. A collision between the main body of CO 0.02--0.02 ($\VLSR\!\simeq\! 100$ \kms ) and the Large Shell ($\VLSR\!\simeq\! 120$ \kms ) may have generated the northwestern protrusion and F2, which broadened the velocity extent of the cloud to $\VLSR\!\simeq\! 140$ \kms . The Small Shell could be one of the bubbles that form the Large Shell, whereas the Emission Cavity is outside its impact. This scenario assumes that the Large Shell has swept a part of CO 0.02--0.02 and its bulk remains unperturbed. The Slit cannot be reproduced in this scenario because its central velocity clearly differs from that of the Large Shell.

\subsection{Formation of HV and UCCs}
The formation of the HV is not included in either the endogenesis or exogenous scenario. Thus, an additional acceleration process is necessary to account for the presence of the HV and UCCs. The absence of expanding features in the HV may indicate acceleration by a nonexplosive process.  

Such an acceleration process has been proposed to interpret the compact high-velocity feature in the W44 molecular cloud~\citep[``Bullet";][]{Yamada17}, where a high-velocity plunge of a point-like massive object is assumed. A magnetohydrodynamic simulation of a plunging black hole into a molecular cloud has successfully reproduced the observed position-velocity behavior of the Bullet \citep{Nomura18}. Another example of gravitationally-kicked gas is CO--0.40--0.22, of which a point-like mass of $10^5$ $M_{\sun}$ was assumed to be responsible for this origin\citep{Oka16}. ALMA observations of HCN--0.009--0.044 and HCN--0.085--0.094 revealed that the gas kinematics in each of these clouds could be reproduced by Keplerian orbit(s) around a point mass of $\sim\! 10^4\,M_{\sun}$~\citep{Takekawa19,Takekawa20}.

The presence of UCCs in the HV suggests gravitational acceleration by a group of point-like massive objects. Because we could not find any counterparts at the positions of UCCs in the mid-infrared images of Spitzer/IRAC~\citep{Stolovy06, Churchwell09} and the X-ray point source catalog by the Chandra X-ray Observatory~\citep{Muno09}, the accelerating source may be a cluster of less (or non-) luminous stars. We suppose that such a less luminous star cluster resides in CO 0.02--0.02, at the position of the HV. The passage of the Large Shell may increase the interaction with stars, generating a number of ultra-compact clumps of high-velocity gas. The HV may be caused by the aggregation of gravitationally kicked gas. 

\subsection{Hybrid Formation Scenario}
As described above, neither the endogenous nor exogenous scenario can reproduce all the observed features in CO 0.02--0.02. Therefore, it is reasonable to expect that both processes are occurring there. Here, we describe the most plausible ``hybrid''-formation scenario for CO 0.02--0.02:  
\begin{enumerate}
\setlength{\itemsep}{0mm}
\setlength{\parsep}{0mm}

\item The filamentary molecular cloud along the F1 direction is the progenitor of CO 0.02--0.02.
\item A huge explosion, which may be responsible for the formation of the Polar Arc, generated the Large Shell.
\item A microburst of star formation occurred at the off-center of the progenitor.
\item A subsequent series of supernova explosions formed the Emission Cavity and Slit.
\item The rim of the Large Shell reached the progenitor cloud to form F2.
\item The passage of the Large Shell at the less luminous star cluster locally accelerated molecular gas to form the HV and UCCs.
\end{enumerate}

The formation scenario described above can reproduce all the observed features in the CO 0.02--0.02 well. The expansion times of $\sim\! 10^{5}$ yr and $\sim\! 10^{4}$ yr for the Large Shell and Slit, and the timescale of $\sim\! 10^{3}$ yr for forming a UCCs-like feature~\citep{Nomura18} also agree with the scenario. The remaining issues are (1) the origin of the progenitor cloud and (2) the nature of the putative star cluster in the HV. Because of the intense tidal field at the location of CO 0.02--0.02, even massive star clusters would dissolve into the background less than $10^6$ yr~\citep{Portegies-Zwart02}. We expect infrared proper motion surveys with high-resolution and high-sensitivity observations to enable us to identify fragmented star clusters, which may be the driving source of the UCCs. In addition, proper motion studies of the UCCs with ALMA will provide clues to unveil the true origin of this peculiar molecular cloud in the CMZ.

\section{Summary}
We have conducted ALMA observations toward the central region of CO 0.02--0.02 in the CO {\it J}=3--2, H$^{13}$CN {\it J}=4--3, H$^{13}$CO$^{+}$ {\it J}=4--3, SiO {\it J}=8--7, and CH$_3$OH {\it J}$_{\it K_a, K_c}$ = 7$_{1, 7}$--6$_{1,6}$ A$^{+}$ lines and the $900$ {\rm $\mu$m} continuum emission. The obtained 1\arcsec\ resolution images revealed the detailed morphology and kinematics of CO 0.02--0.02. The main conclusions are summarized as follows:
\begin{enumerate}
\setlength{\itemsep}{0mm}
\setlength{\parsep}{0mm}
\item The images of dense gas probes (H$^{13}$CN, SiO, and CH$_3$OH) are dominated by the two filaments (named Filaments), which stretch along the northeast to southwest direction. These Filaments, appearing at $\VLSR\!=\!60\mbox{--}120$ \kms , may be the spine of CO 0.02--0.02.  
\item The brightest continuum core (BCC) was found at the southwestern edge of the Filaments. All the observed molecular lines were detected from the BCC. Its velocity dispersion ($\sigma_{V}\!=\! 6.3$ \kms ) is one order of magnitude higher than that expected from the $S$-$\sigma_{V}$ relation of the CMZ clouds at the size of the BCC ($S\!=\! 0.05$ pc).
\item Less dense gas probe (CO) images exhibited three characteristic features of F1, F2, and HV, which appear at $\VLSR\!=\! 0\mbox{--}120$ \kms , $\VLSR\!=\! 100\mbox{--}140$ \kms , and $\VLSR\!\geq\! 140$ \kms , respectively.  
\item F1 is a northeast-southwest striped feature that represents the main body of CO 0.02--0.02.  The Filaments align along F1. An emission hole, which may be an expanding feature, aligned along northeast-southwest (the Slit) was identified in the velocity extent of F1. The central velocity of the Slit is similar to that of the southwestern Emission Cavity.  
\item F2 is a southeast-northwest striped feature appearing at a positive high-velocity portion of CO 0.02--0.02. The direction of the F2 stripes is similar to that of the western edge of the $20\,\mbox{pc}\times 13\,\mbox{pc}$ expanding shell (the Large Shell), which was identified in the large-scale CO images.
\item The HV is a high-velocity clump that arises from F2, being located on the edge of the Large Shell. Eight ultra-compact clumps (UCCs) were identified at the positive high-velocity end of CO emission. The absence of expanding features in the HV is indicative of the acceleration by a nonexplosive process.  
\item We proposed a ``hybrid'' formation scenario for CO 0.02--0.02; internal explosions of supernovae (the Slit and Emission Cavity), external perturbation by the Large Shell, and gravitational acceleration by a less luminous star cluster (the HV and UCCs).
\end{enumerate}

\begin{acknowledgments}
YI was supported by the ALMA Japan Research Grant of National Astronomical Observatory of Japan (NAOJ) ALMA Project, NAOJ-ALMA-270. Data analysis was in part carried out on the Multi-wavelength Data Analysis System operated by the Astronomy Data Center (ADC), NAOJ. This paper makes use of the following ALMA data: ADS/JAO.ALMA\#2019.1.01763.S. ALMA is a partnership of ESO (representing its member states), NSF (USA) and NINS (Japan), together with NRC (Canada), MOST and ASIAA (Taiwan), and KASI (Republic of Korea), in cooperation with the Republic of Chile. The Joint ALMA Observatory is operated by ESO, AUI/NRAO and NAOJ.
\end{acknowledgments}

\vspace{5mm}
\facilities{ALMA.}

\software{CASA \citep{McMullin07,CASA22}, Astropy \citep{astropy13,astropy18}, APLpy \citep{Robitaille12}.}

\bibliography{references}{}
\bibliographystyle{aasjournal}

\end{document}